\documentstyle[aps,prl,epsf,floats,multicol,amssymb,tighten]{revtex}
\begin{document}
\newcommand{\base}{}
\newcommand{\RA}{\rangle}
\newcommand{\LA}{\langle}
\newcommand{\RR}{\rangle\rangle}
\newcommand{\LL}{\langle\langle}
\newcommand{\nonb}{\nonumber}

\newenvironment{tab}[1]
{\begin{tabular}{|#1|}\hline}
{\hline\end{tabular}}

\newcommand{\fig}[2]{\epsfxsize=#1\epsfbox{#2}} \reversemarginpar 
\bibliographystyle{prsty}

\title{Unconventional antiferromagnetic correlations of the\\
doped Haldane gap system Y$_2$BaNi$_{1-x}$Zn$_x$O$_5$}

\author{
V. Villar$^{(1)}$
,
R. M\'elin$^{(1)}$\thanks{melin@polycnrs-gre.fr}
,
C. Paulsen$^{(1)}$
,
J. Souletie$^{(1)}$
,
E. Janod$^{(2)}$
,
C. Payen$^{(2)}$
}

\address{$^{(1)}$ Centre de Recherches sur les Tr\`es Basses
Temp\'eratures (CRTBT)\thanks{U.P.R. 5001 du CNRS,
Laboratoire conventionn\'e avec l'Universit\'e Joseph Fourier},\\
CNRS BP 166X, 38042 Grenoble Cedex, France}
\address{$^{(2)}$ Institut des Mat\'eriaux Jean Rouxel,
Universit\'e de Nantes-CNRS, 44322 Nantes Cedex 3, France}
\date{\today}
\maketitle

\begin{abstract}
We make a new proposal to describe
the very low temperature susceptibility
of the doped Haldane gap compound
Y$_2$BaNi$_{1-x}$Zn$_x$O$_5$.
We propose a new mean field model relevant for
this compound.
The ground state of this mean field model
is unconventional because
antiferromagnetism coexists with random
dimers.
We  present new susceptibility experiments at
very low temperature.
We obtain a Curie-Weiss
susceptibility $\chi(T) \sim C / (\Theta+T)$
as expected for antiferromagnetic correlations but
we do not obtain a direct signature
of antiferromagnetic long range order.
We explain how to obtain the ``impurity''
susceptibility $\chi_{\rm imp}(T)$
by subtracting the Haldane
gap contribution to the total
susceptibility. In the temperature
range [1~K, 300~K] the experimental 
data are well fitted by
$T \chi_{\rm imp}(T) = 
C_{\rm imp} \left( 1 + T_{\rm imp}/T
\right)^{-\gamma}$.
In the
temperature range [100~mK, 1~K] the 
experimental data are well
fitted by
$T \chi_{\rm imp}(T) = A \ln{(T/T_c)}$,
where $T_c$ increases with $x$. This fit
suggests the existence of a finite
N\'eel temperature which is however too small
to be probed directly in our experiments.
We also obtain a maximum in the temperature
dependence of the ac-susceptibility
$\chi'(T)$ which suggests the existence
of antiferromagnetic correlations
at very low temperature.
\end{abstract}

\widetext

\section{Introduction}

Disordered low dimensional spin systems are currently
the focus of both experimental and
theoretical interest.
A very powerful theoretical technique
used to describe these systems consists in 
iterating a real space renormalization group (RG) in
which high energy degrees of freedom are progressively
frozen out. The cluster RG was initially proposed
by Dasgupta an Ma in the early 80's~\cite{Dasgupta}
and applied soon after by Bhatt and Lee to
a three dimensional (3D) model 
intended to describe Si:P~\cite{Bhatt}.
In a series of articles, Fisher applied
the cluster RG to 
disordered 1D antiferromagnets, including
the random Ising chain in a transverse magnetic
field and 
the random antiferromagnetic spin-$1/2$
chain~\cite{Fisher}. In these models, the cluster RG is so powerful
that it allows to obtain {\sl exact results}
regarding two aspects of the problem:
(i) the correlation functions associated
with the approach of a random singlet fixed point;
and (ii) the approach of a quantum
critical point.
It is fair to say that
there exists now a detailed theoretical understanding
of all random spin models
in one dimension~\cite{dimer,spin1-a,spin1-b,Westerberg},
and even in higher dimensions~\cite{Motrunich}.
Some models do not have a quantum critical
point~\cite{Fisher,Westerberg} while
other models have a zero temperature phase transition
which is controlled by the strength of
disorder~\cite{Fisher,dimer,spin1-a,spin1-b}.
In particular, weak disorder is
irrelevant in the random spin-$1$
chain but there is a quantum critical
point at a critical disorder $d_c$.
For $d > d_c$, the random spin-$1$
chain behaves like a random
singlet~\cite{spin1-a,spin1-b},
namely, the strength of disorder
increases indefinitely as the 
temperature is scaled down, which
is known as an ``infinite
randomness'' behavior~\cite{Motrunich}.

On the experimental side, it has
become possible to
fabricate low dimensional oxides in the
last five years, and dope
them in a well controlled fashion. As a 
consequence, these systems give the
unique opportunity of exploring the {\sl introduction
of disorder in a spin gap state}.
It would be erroneous however to think 
that there exists a straightforward
relation between these
experiments on quasi one dimensional oxides and the
1D models that have been widely studied by
theoreticians in which the disorder is introduced
in the form of random bonds.
In fact, the
1D models with random bonds are not directly
applicable to experiments. The reason
is that the relevant models should incorporate
the following {\sl realistic
constraints}:
\begin{itemize}
\item[(i)] Doping in quasi one dimensional oxides
is introduced by substitutions, and not in the
form of random exchanges.
The usual random bond models are not the
most relevant ones. 
\item[(ii)] The temperature is finite in the experiments,
even though often very small.
\item[(iii)] The realistic systems are
not 1D.
Interchain couplings become relevant at
low temperature and can change the physics
drastically.
\end{itemize}
The problem is then to determine to
what extent realistic models incorporating
these three constraints still
behave like the original 1D
random bond models.

One of the recently discovered low
dimensional oxides is CuGeO$_3$~\cite{SP}
which has a spin-Peierls transition
at $T_{SP}\simeq 14$~K below which
the CuO$_2$ chains dimerize with
the appearance of a gap in the
spin excitation spectrum.
Soon after the discovery experimentalists
began to study the effects of various
substitutions on this inorganic compound.
For
instance, it is possible to substitute
a small fraction of
the Cu sites (being a spin-$1/2$ ion)
by Ni (being a spin-$1$ ion)~\cite{Ni}
or Co (being a spin-$3/2$ ion)~\cite{Co}.
It is possible to substitute
Cu by the non magnetic ions
Zn~\cite{Zn,Hase,Martin} or Mg~\cite{Mg}.
It is also possible to substitute
some Ge sites (being outside the
CuO$_2$ chains) with Si~\cite{Si}.
The general feature emerging from
the detailed experimental studies
of the various
substitutions is the existence of
antiferromagnetism at low temperature.
It has even been shown experimentally
by Manabe {\sl et al.} that
at low doping concentrations, the N\'eel temperature
behaves like $\ln{T_N} \sim 1/x$~\cite{Manabe98},
therefore suggesting that there is no
critical concentration associated
with the onset of antiferromagnetism.

\begin{figure}[thb]
\centerline{\fig{8cm}{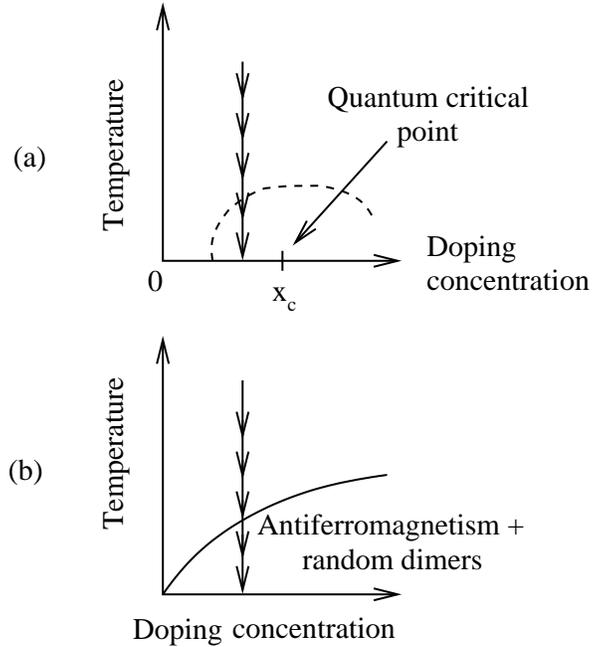}} 
\medskip
\caption{Two possible scenarii for
Y$_2$BaNi$_{1-x}$Zn$_x$O$_5$. In (a)
there are no interchain couplings
($J_\perp=0$) and the physics is governed by
the approach to the quantum critical point of the
disordered Haldane gap chain. A power-law susceptibility
is expected in this situation at low temperature
(Griffiths behavior).
In (b) interchain
couplings play a relevant role and generate
an unconventional phase in which antiferromagnetism
coexists with random dimers. The approach
to the N\'eel transition is controlled by the
establishment of antiferromagnetic correlations.
The purpose of the article is to show that
the situation (b) can apply to Y$_2$BaNi$_{1-x}$Zn$_x$O$_5$.
} 
\label{fig:phases}
\end{figure}

The Haldane gap is another example of a
spin gap in a low
dimensional antiferromagnet~\cite{Haldane}.
Two inorganic Haldane gap
antiferromagnets have been discovered
recently: PbNi$_2$V$_2$O$_8$
which has a spin gap
$\Delta \simeq 28$~K~\cite{Pb}, and
Y$_2$BaNiO$_5$ which has a spin gap
$\Delta \simeq 100$~K~\cite{ori-Y}.
The substitution of the spin-$1$ Ni sites of
PbNi$_2$V$_2$O$_8$ with Mg (a non magnetic ion)
generates long range antiferromagnetism at low
temperature.
The situation with Y$_2$BaNiO$_5$ is not
so well established. Previous works
have failed to find long range antiferromagnetic
order
at low temperature~\cite{Batlogg,DiTusa,Kojima,Payen00}.
In particular $\mu$SR experiments at
$50$~mK in Ref.~\cite{Kojima} have reported
a paramagnetic relaxation with a Mg
doping of $1.7\%$ and $4.1\%$.
We report here a study of the effect of Zn
substitutions in Y$_2$BaNiO$_5$. In agreement
with Ref.~\cite{Kojima}, we
do not find
antiferromagnetic long range order
down to $100$~mK. But from the analysis
of the experimental ac- and dc-susceptibility we deduce
the existence of 3D antiferromagnetic correlations
which rule out the
``1D quantum criticality'' scenario represented
on Fig.~\ref{fig:phases}-(a).

There have been several attempts to find
a theoretical description of doped spin-Peierls
systems. Fukuyama, Tanimoto and Saito
have made the first
proposal~\cite{Fuku}. 
Another route has been followed by
Fabrizio, M\'elin and
Souletie. In a series of article~\cite{Melin-SP,Melin-spin1,Melin00},
these authors have
provided a detailed
scenario for disordered antiferromagnetism.
Appealing aspects of their
approach is that the doped spin-Peierls and
the doped Haldane gap systems can be
described in the same framework~\cite{Melin00},
and the model appears to be fully compatible
with experiments.
The relevant physics is determined by
comparing two energy scales:
\begin{itemize}
\item[(i)] The coherence temperature $T^* = \Delta
\exp{[-1/(x \xi)]}$ which controls the
formation of singlet correlations.
\item[(ii)] The Stoner
temperature $T_{\rm Stoner} = J_{\perp}
x \xi$ which controls the formation of
3D antiferromagnetic correlations.
\end{itemize}
In these expressions,
$x$ is the doping concentration,
$\Delta$ is the spin gap,
$\xi$ is the correlation length of the
pure gaped system, and $J_\perp$ is
the interchain coupling. 
The specificity of Y$_2$BaNi$_{1-x}$Zn$_x$O$_5$
is that $T^*>T_{\rm Stoner}$  while
$T^* \ll T_{\rm Stoner}$ in Cu$_{1-x}$Zn$_x$GeO$_3$
and Pb(Ni$_{1-x}$Mg$_x$)V$_2$O$_8$ (see Refs.~\cite{Melin-SP,Melin00}).
As a consequence in Y$_2$BaNi$_{1-x}$Zn$_x$O$_5$,
3D antiferromagnetic
correlations coexist with non magnetic
random dimer correlations
(see Fig.~\ref{fig:phases}--(b)). 
Based on the calculation 
of the exchange coupling two spin-$1/2$ moments
associated to the same impurity presented
in Ref.~\cite{Chemist}, we propose
a mean field model relevant for 
Y$_2$BaNi$_{1-x}$Zn$_x$O$_5$. From this
mean field model we can obtain new insights in the
nature of the low temperature phase, and
compare the model to very low
temperature susceptibility experiments.

The article is organized as follows.
The model is presented in section~\ref{sec:themodel}.
Section~\ref{sec:coexistence} is devoted to
calculate the susceptibility and discuss
the nature of the low temperature phase.
In section~\ref{sec:experiments} 
we present and analyze the very low temperature
susceptibility experiments.
Concluding remarks are given in section~\ref{sec:conclusion}.

\section{The model}
\label{sec:themodel}

\subsection{Defects in spin-Peierls and
Haldane gap systems}
\begin{figure}[thb]
\centerline{\fig{9cm}{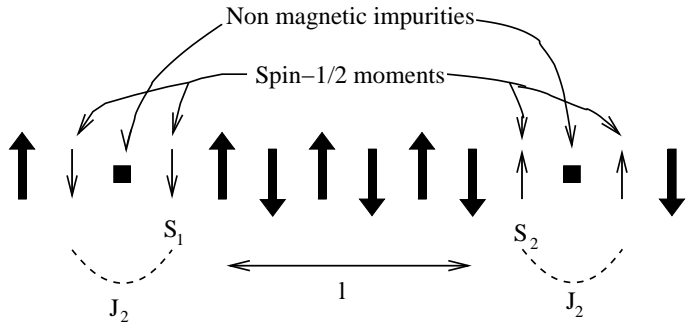}} 
\medskip
\caption{Schematic representation of two non
magnetic impurities inserted in a spin-$1$ chain.
The thick arrows are the spin-$1$ moments. The
thin arrows are the ``edge'' spin-$1/2$ moments. The
exchange coupling the
two spins ${\bf S}_1$ and ${\bf S}_2$
is given by
(\ref{eq:H}) and (\ref{eq:J}). We have
represented an even segment.
The exchange between
${\bf S_1}$ and ${\bf S}_2$ is therefore
antiferromagnetic.} 
\label{fig:schema}
\end{figure}

\begin{figure}[thb]
\centerline{\fig{9cm}{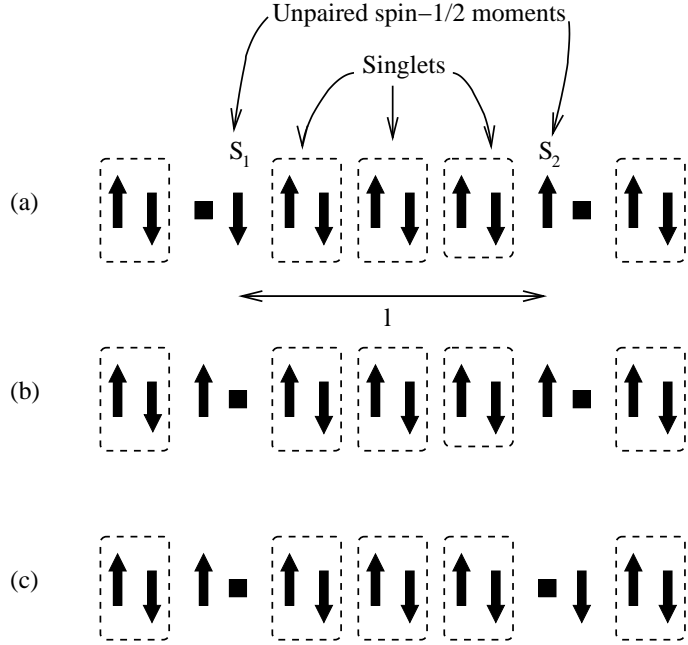}} 
\medskip
\caption{Schematic representation of two non
magnetic impurities inserted in a spin-Peierls chain.
(a) corresponds an even segment with
two spin-$1/2$ moments. In this case
the exchange (\ref{eq:H}), (\ref{eq:J})
between ${\bf S}_1$ and ${\bf S}_2$
is the same as for the Haldane gap chain.
(b) corresponds to
an odd segment with one spin-$1/2$ moment.
(c) corresponds to an even segment with
no spin-$1/2$ moment.
} 
\label{fig:SP}
\end{figure}
The model that should be used to describe doped
Haldane gap compounds is well established.
The introduction of a non-magnetic impurity
in a spin-$1$ chain generates a pair of
``edge'' spin-$1/2$
moments on either side of the
impurity~\cite{Kennedy,Hagiwara90,Sorensen,Yamamoto}
(see Fig.~\ref{fig:schema}).
The spin-$1/2$
moments associated with two neighboring impurities
interact with the Hamiltonian
\begin{equation}
\label{eq:H}
{\cal H}=J(l) {\bf S}_1 . {\bf S}_2
,
\end{equation}
where the exchange is ferromagnetic for odd segments
(${\bf S}_1$ and ${\bf S}_2$ are in the same
sublattice),
and antiferromagnetic for even segments
(${\bf S}_1$ and ${\bf S}_2$ are in a different
sublattice):
\begin{equation}
\label{eq:J}
J(l) = (-)^l \Delta \exp{\left(- \frac{l}{\xi} \right)}
,
\end{equation}
where $\xi$ is the correlation length associated to
the Haldane gap. 
We note $J_2$ the ferromagnetic exchange
coupling between two spins associated with the same
impurity. Quantum chemistry calculations
indicate that $J_2$ is smaller
than the interchain coupling $J_\perp$
in Y$_2$BaNi$_{1-x}$Zn$_x$O$_5$~\cite{Chemist}.
The authors of Ref.~\cite{Chemist}
have obtained
$J_\perp \simeq 0.2 - 0.9$~K
and $J_2 <0.1$~K.
This means that there is no
temperature range in which the effective
low energy model can be considered as
one dimensional. On the contrary,
when $T$ is above $J_2 \sim J_\perp$
the system can be represented by a
model having $J_2=J_\perp=0$.
When $T$ is below $J_2 \sim J_\perp$ the
system develops directly three-dimensional
correlations. As a consequence we will
use a model with $J_2=0$ but with finite
interchain correlations that will be treated
in mean field. We note that the role
of interchain correlations was already
pointed out in Ref.~\cite{Payen00}. On the
basis of DMRG calculations for open segments,
the authors of Ref.~\cite{Payen00} could
obtain a good agreement between the model
and the experiments on Y$_2$BaNi$_{1-x}$Zn$_x$O$_5$
above 4~K. The deviations occurring in the
temperature range [2~K, 4~K] have been
attributed to interchain correlations which
play a central role in the model developed
here.

It will be fruitful to make a qualitative comparison
with doped spin-Peierls
systems, which are closely related to the doped
Haldane gap systems~\cite{Melin-SP}. 
Non magnetic impurities introduced in
a spin-Peierls system generate unpaired
spin-$1/2$ moments (see Fig.~\ref{fig:SP}).
Depending
on the parity of the length of the segment
and on the dimerization pattern, there can
be zero, one or two spin-$1/2$ moments.
If there are two spin-$1/2$ moments
(see Fig.~\ref{fig:SP}-(a))
the exchange
(\ref{eq:H}) and (\ref{eq:J}) is the same as
for the Haldane gap chain. To discuss
the qualitative physics, we will make the
approximation of including only the segments
on Fig.~\ref{fig:SP}-(a) that give rise to
two spin-$1/2$ moments and neglect the
other segments.

\subsection{Role of interchain interactions}

Interchain interactions can be 
described by introducing a correlation
length $\xi_\perp$ in the transverse direction.
In two dimensions, the
effective low energy Hamiltonian
is a sum over all pairs of spins~\cite{Melin-SP}:
\begin{equation}
\label{eq:H2D}
{\cal H} = \sum_{\langle i,j \rangle}
J_{i,j} {\bf S}_i.{\bf S}_j
,
\end{equation}
where the exchange between the spin
${\bf S}_i$ at coordinates $(x_i,y_i)$
and the spin ${\bf S}_j$ at coordinates
$(x_j,y_j)$ is given by
\begin{equation}
\label{eq:J2D}
J_{i,j} =
(-)^{r_x+r_y} \Delta
\exp{\left( - \sqrt{
\left( \frac{r_x}{\xi} \right)^2
+ \left( \frac{r_y}{\xi_\perp} \right)^2}
\right)}
,
\end{equation}
where $r_x=x_i-x_j$ and $r_y=y_i-y_j$. It has
been shown from the RG approach in 2D
that the Hamiltonian defined
by (\ref{eq:H2D}) and (\ref{eq:J2D}) in two
dimensions has a
{\sl finite randomness} behavior~\cite{Melin-SP}.
This indicates the existence of
long range antiferromagnetism.
In the present work, we do not address directly
the model defined by~(\ref{eq:H2D}) and~(\ref{eq:J2D})
in 2D. Instead we
propose that Y$_2$BaNi$_{1-x}$Zn$_x$O$_5$ can be
described by a mean field model
in which we introduce a coupling
to a molecular field $-h_s \sum_i S_i^z$.
The value of the molecular field is equal
to the strength of antiferromagnetic correlations,
which can be obtained from the susceptibility
experiments in section~\ref{sec:experiments}.
Since we use a mean field model,
we cannot make the distinction between
``antiferromagnetic long range order''
and ``antiferromagnetic correlations
with a finite (but possibly large) correlation length''.
For the sake of
simplicity we assume that the staggered
molecular field is uniform. But our approach can be
extended to incorporate a distribution of
molecular fields. The distribution of molecular
fields can be calculated in a self-consistent
way by imposing that the exchange field distribution
is identical to the staggered moment distribution,
up to a proportionality factor related to the
strength of interchain correlations.

\subsection{Summary of the article}
\begin{figure}[thb]
\centerline{\fig{8cm}{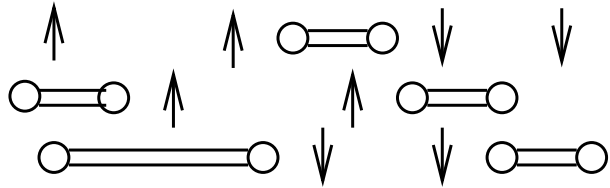}} 
\medskip
\caption{Schematic representation of the ground
  state obtained from the mean field approach.
  The even segments are coupled
  into non magnetic random dimers.
  The other spins give rise to antiferromagnetic
  long range order.} 
\label{fig:coex}
\end{figure}
\label{sec:sum}
Our strategy will be to solve the cluster
RG in the presence of a molecular field
for a  model having
$J_2=0$ (see Fig.~\ref{fig:schema}).
Because this model reduces to a sum
of independent two-spin Hamiltonians, we
call this model a ``two-spin model''. The
Hamiltonian is given in Appendix~\ref{app:exact}
(see Eqs.~\ref{eq:H-odd},~\ref{eq:H-even}).

The mean  field model shows that
antiferromagnetism coexists
with random dimers at low temperature (see
section~\ref{sec:coexistence}) which are
due to the even segments coupled
antiferromagnetically. This coexistence
can be understood on a simple basis: at low temperature
the even segments are coupled into
non magnetic dimers while the odd
segments are coupled into spin-$1$ objects.
Interchain interactions generate antiferromagnetic
correlations among the spin-$1$ moments.
The spin-$1$
moments can eventually order
antiferromagnetically (see Fig.~\ref{fig:coex}).

\section{Nature of the low temperature phase:
coexistence between antiferromagnetism and
random dimers}
\label{sec:coexistence}

\subsection{Susceptibility of the two-spin model}
\label{sec:exact}
\begin{figure}[thb]
\centerline{\fig{9cm}{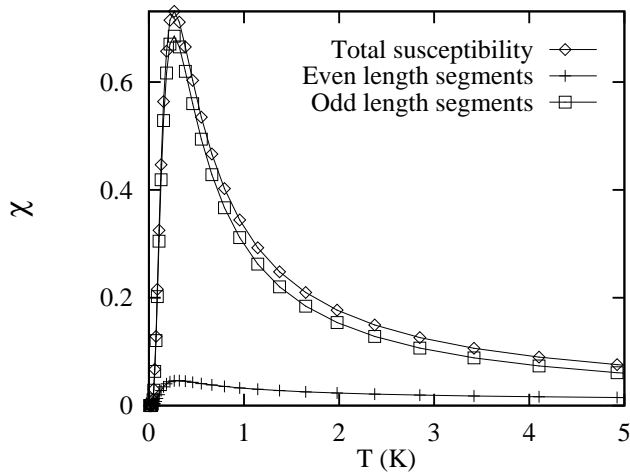}} 
\medskip
\caption{Temperature dependence of the susceptibility
  of the two-spin model, with the parameters relevant
  for Y$_2$BaNiO$_5$: $\xi=6$, $\Delta=100$~K,
  $h_s=0.3$~K, $x=8\%$. We have also shown separately the
  contribution of even segments (coupled
  antiferromagnetically) and odd
  segments (coupled ferromagnetically).} 
\label{fig:chi-2spin}
\end{figure}

\begin{figure}[thb]
\centerline{\fig{9cm}{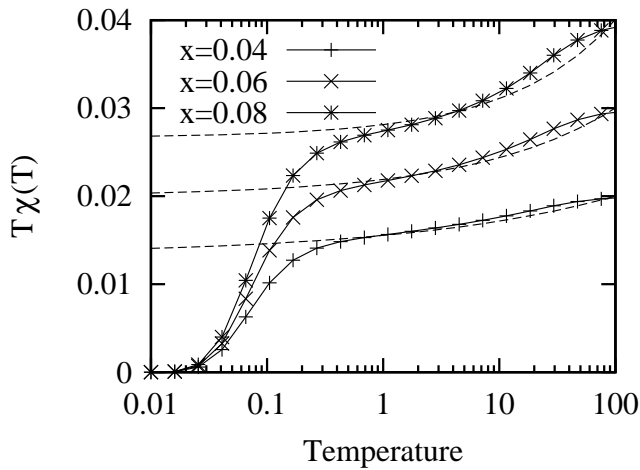}} 
\medskip
\caption{Temperature dependence of $T \chi(T)$
  for the two-spin
  model coupled to a molecular field. The parameters
  are identical to Fig.~\ref{fig:chi-2spin}. 
  For each doping concentration we have
  shown by a dashed line
  the temperature dependence of the
  RG susceptibility (\ref{eq:chi-tot-RG})
  which does not contain a coupling to
  the molecular field.} 
\label{fig:chiT-th}
\end{figure}

The exact treatment of the two-spin model
is briefly given in Appendix~\ref{app:exact}.
The total susceptibility is obtained by averaging
Eq.~\ref{eq:magne} over all possible 
realizations of disorder, {\sl i.e.} over all
possible bond lengths:
$$
\LL \chi(T) \RR = \sum_{l=1}^{+ \infty}
{\cal P}(l) \chi(l,T)
.
$$
The bond length distribution
is ${\cal P}(l)=x (1-x)^{l-1} \simeq
x \exp{(-xl)}$,
with $x$ the impurity concentration.
The temperature dependence of the susceptibility
is shown on Fig.~\ref{fig:chi-2spin}. A maximum occurs
in the susceptibility
when the temperature is of the
order of interchain interactions,
below which the susceptibility decreases to zero.
As it can be seen in Fig.~\ref{fig:chi-2spin},
the contribution of even segments
(being coupled antiferromagnetically)
is much smaller than the contribution of
odd segments (being coupled ferromagnetically).
The reason is that
even segments tend to form dimers
which do not couple to a uniform magnetic
field. Both types of segments couple to the
staggered molecular field, which is why the susceptibility
on Fig.~\ref{fig:chi-2spin} has a maximum when
$T \simeq h_s$ and decreases to zero at lower temperature.

The temperature dependence of $T \chi(T)$ is shown
on Fig.~\ref{fig:chiT-th} for three doping concentrations
corresponding to the experiments in section~\ref{sec:experiments}.
At high temperature we recover the Curie constant
of free spin-$1/2$ moments with a concentration
$2x$:
$$
\left[ T \chi(T) \right]_{T=\Delta}
= 2 x \frac{S(S+1)}{3} = {x \over 2}
.
$$
At low temperature, $T \chi(T)$ is
exponentially small because of the coupling
to the staggered molecular field.

\subsection{Renormalization of the two-spin model}
\label{sec:RG-2spin}
\begin{figure}[thb]
\centerline{\fig{9cm}{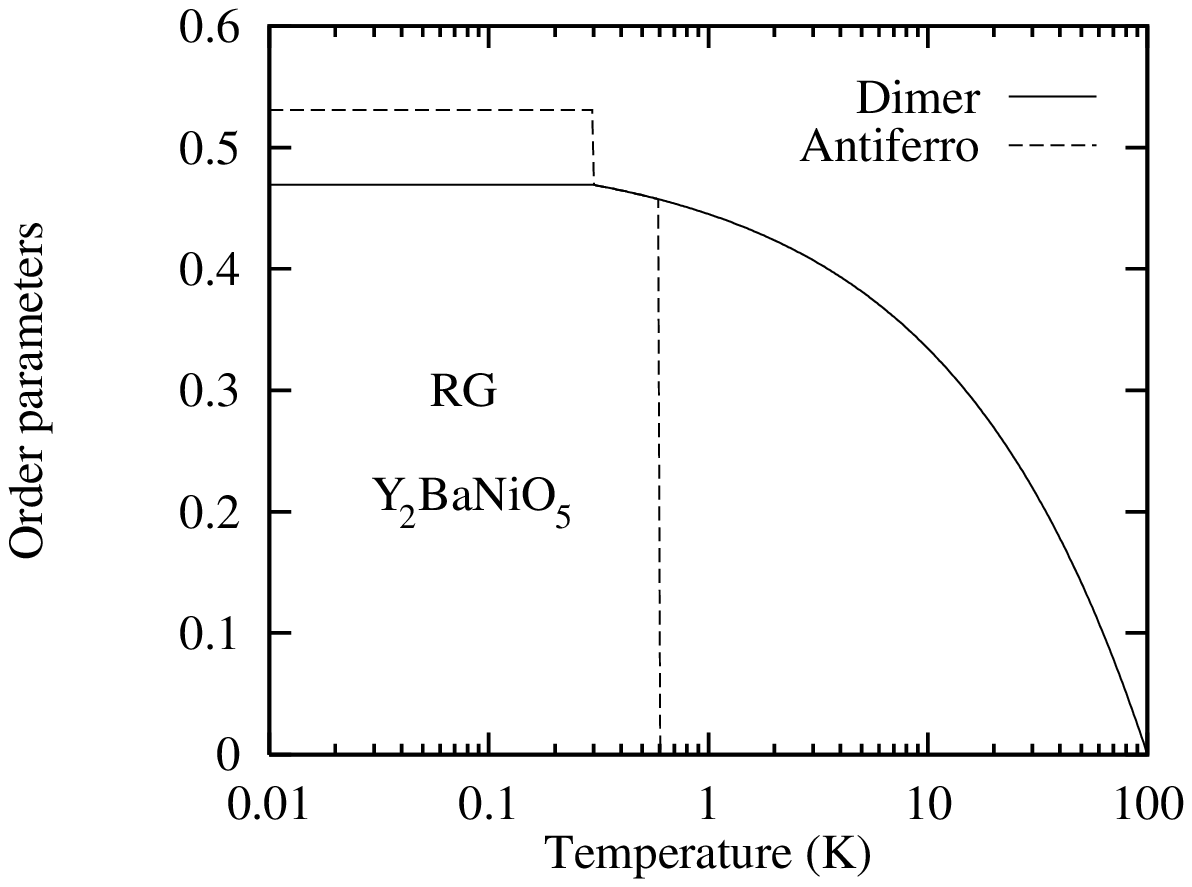}} 
\medskip
\caption{Temperature dependence of the
two order parameters $\varphi_{\rm AF}(T)$
and $\varphi_{\rm dim}(T)$ with the parameters
relevant for Y$_2$BaNiO$_5$ (see Fig.~\ref{fig:chi-2spin}).
The energy scale corresponding to random
dimer formation is larger than the energy
scale at which antiferromagnetism appears.
The calculation corresponds to the RG procedure
presented in section~\ref{sec:RG-2spin}.
The parameters are given on Fig.~\ref{fig:chi-2spin}.}
\label{fig:rg-Y2}
\end{figure}
\begin{figure}[thb]
\centerline{\fig{9cm}{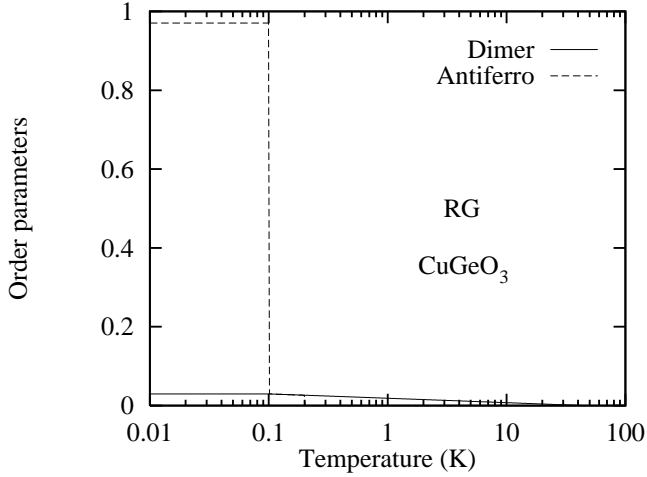}} 
\medskip
\caption{Temperature dependence of the
two order parameters $\varphi_{\rm AF}(T)$
and $\varphi_{\rm dim}(T)$ with the parameters
relevant for CuGeO$_3$ (see Fig.~\ref{fig:orderCu}).
Dimer formation plays little role here.
The calculation corresponds to the RG procedure
presented in section~\ref{sec:RG-2spin}.
The parameters are $\Delta=44.7$~K,
$\xi=10$, $h_s=0.03$ and $x=0.001$.
The values of $h_s$ and $x$ are deduced from
the low doping experiments in Ref.~\protect\cite{Manabe98}.
}
\label{fig:rg-Cu}
\end{figure}
\begin{figure}[thb]
\centerline{\fig{9cm}{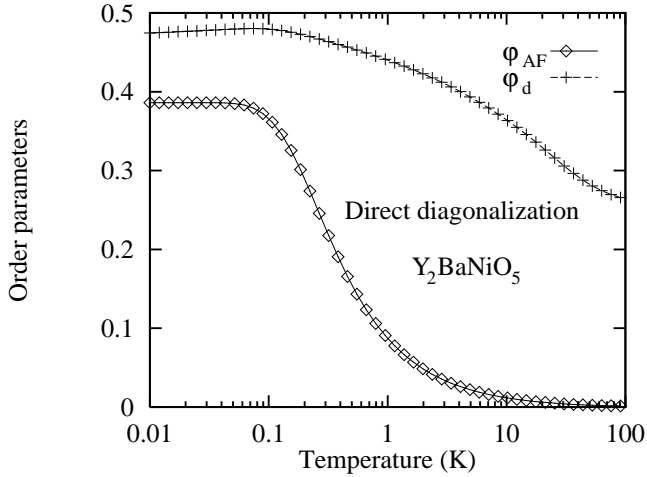}} 
\medskip
\caption{Antiferromagnetic and dimer order parameters
with the parameters relevant for Y$_2$BaNiO$_5$ (see
Fig.~\ref{fig:chi-2spin}). The calculation corresponds
to a direct diagonalization of the two-spin
Hamiltonian, subject to a temperature-independent
staggered molecular field.
}
\label{fig:orderY}
\end{figure}
\begin{figure}[thb]
\centerline{\fig{9cm}{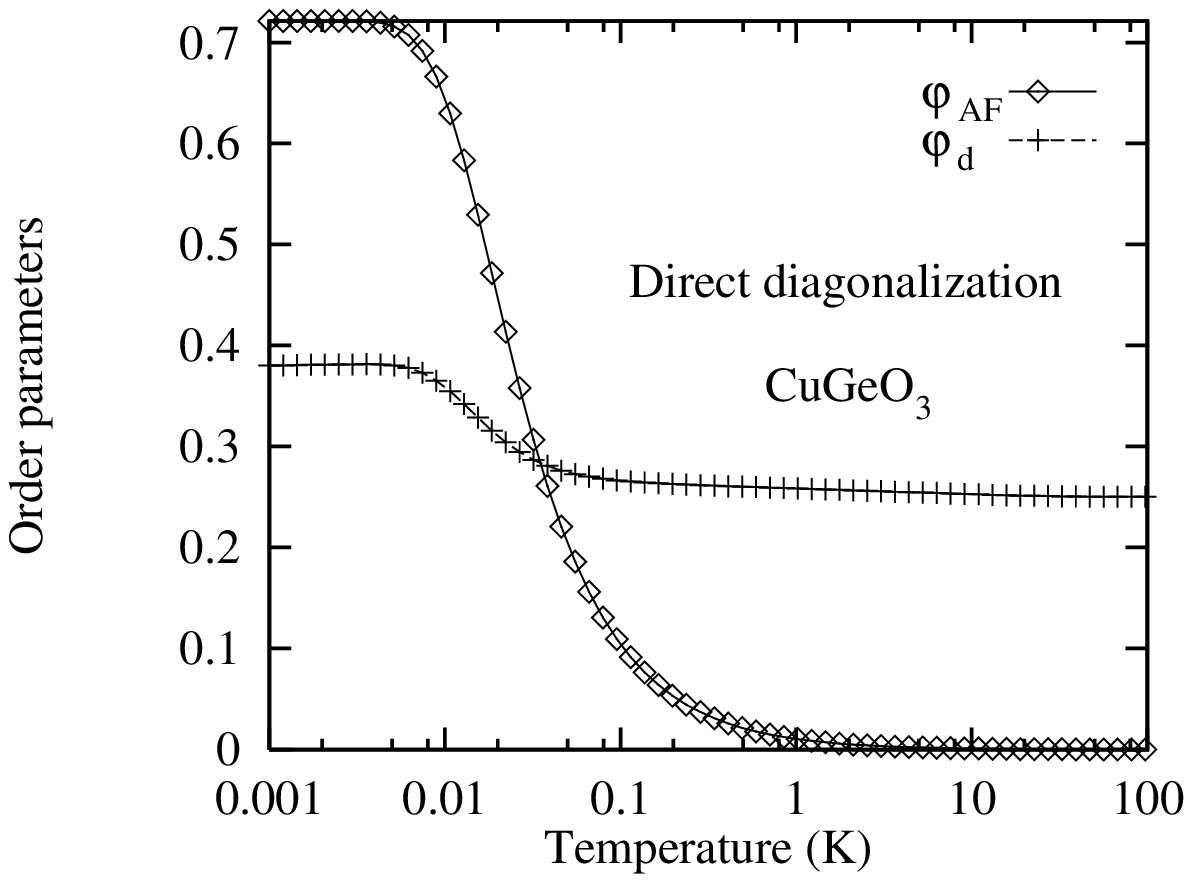}} 
\medskip
\caption{Antiferromagnetic and dimer order parameters
with the parameters relevant for CuGeO$_3$.
The parameters are shown on Fig.~\ref{fig:rg-Cu}.
}
\label{fig:orderCu}
\end{figure}

Now we implement RG transformations for the
two-spin model coupled to a molecular field
$h_s$ and having $J_2=0$.
At low temperature ($T<h_s$) the spins are either
frozen into dimers, or align antiferromagnetically.
The coexistence between the two orders
is therefore very natural in this
framework.
The detailed
calculation in presented in Appendix~\ref{app:RG}.

\subsubsection{Order parameters}
The temperature dependence of the two order parameters 
is shown in Fig~\ref{fig:rg-Y2} 
for Y$_2$BaNiO$_5$ and in Fig.~\ref{fig:rg-Cu}
for CuGeO$_3$. It is clear that dimer
formation is not important in CuGeO$_3$ while
it plays a crucial role in Y$_2$BaNiO$_5$.
To check that this coexistence is not an
artifact of the cluster RG, we have made
a similar analysis with the direct
diagonalization of the two-spin
model (see Appendix~\ref{app:exact}).
Considering and eigenstate of the form
$x |+,- \rangle + y |-,+ \rangle$,
we define
\begin{eqnarray}
\varphi_{\rm AF} &=& x^2 - y^2\\
\varphi_{\rm dim} &=& {1 \over 2} (x-y)^2\
\end{eqnarray}
as the order parameters associated
to antiferromagnetism and dimer
formation. 
In the presence of a pure dimer
ordering, one has
$x=-y=1/\sqrt{2}$, leading to
$\varphi_{\rm AF}=0$ and
$\varphi_{\rm dim}=1$.
In the presence of a pure
antiferromagnetic ordering, one has
$x=1$ and $y=0$, leading to
$\varphi_{\rm AF}=1$ and
$\varphi_{\rm dim}=1/2$.
The qualitative
behavior is similar to the RG
treatment (see Figs.~\ref{fig:orderY}
and~\ref{fig:orderCu}), therefore
supporting our proposal that random
dimers coexist with antiferromagnetism
in the mean field model relevant
for Y$_2$BaNiO$_5$.

\subsubsection{RG Susceptibility}
\label{sec:RG-chi}
The RG susceptibility in the temperature
range $2 h_s<T<\Delta$ is obtained as
the sum of the contribution of the
spin-$1/2$ and spin-$1$ moments that
have not been decimated.
We note $n_S(T)$ the density of
spin-$S$ moments (number of spin-$S$ moments
per unit length).
The total susceptibility reads
\begin{equation}
\chi(T) = \sum_S n_S(T) \frac{S(S+1)}{3 T}
.
\end{equation}
As far as even segments are concerned,
we find (see Appendix~\ref{app:RG})
$
n_{1/2}^{({\rm even})}(T) =
x \left( T/\Delta \right)^{x \xi}
$,
from what we obtain the susceptibility of
even segments:
\begin{equation}
\chi_{\rm even}(T) = \frac{x}{4 T}
\left( \frac{T}{\Delta} \right)^{x \xi}
.\label{eq:chi-A}
\end{equation}
As far as odd segments are concerned,
we obtain from Appendix~\ref{app:RG}:
$n_{1/2}^{({\rm odd})}(T) = x \left( T/\Delta
\right)^{x \xi}$, and
$n_1^{({\rm odd})}(T) = {x \over 2} \left[
1 - \left( T/\Delta \right)^{x \xi}
\right]
$,
from what we deduce
\begin{eqnarray}
\label{eq:chi-C}
\chi_{\rm odd,1/2}(T) &=& \frac{x}{4 T}
\left( \frac{T}{\Delta} \right)^{x \xi}
\label{eq:chi-B}\\
\chi_{\rm odd,1}(T) &=& \frac{x}{3 T}
\left[ 1 - \left( \frac{T}{\Delta} \right)^{x \xi}
\right]
.
\end{eqnarray}
The total susceptibility is
$\chi_{\rm tot}(T) = \chi_{\rm even}(T) +
\chi_{\rm odd,1/2}(T) +
\chi_{\rm odd,1}(T)$:
\begin{equation}
\label{eq:chi-tot-RG}
\chi_{\rm tot}(T) = {x \over 3 T}
\left[ 1 + {1 \over 2}
\left( {T \over \Delta} \right)^{x \xi}
\right]
,
\end{equation}
where the temperature $T$
is such that $2 h_s < T < \Delta$.

We have presented on Fig.~\ref{fig:chiT-th}
a comparison between the direct diagonalizations and
the RG susceptibility of the two-spin model.
We see from this figure that the value
of $T \chi(T)$ obtained from the cluster-RG
is very close to the exact diagonalizations
in the temperature range $2 h_s<T<\Delta$.
We can make the following remarks:
\begin{itemize}
\item[(i)] In the absence of the staggered
molecular field ({\sl i.e.} in the absence
of interchain couplings) $T \chi(T)$ tends
to a constant when $T \rightarrow 0$.
This constant is equal to the Curie constant
of spin-$1$ moments with a concentration $x/2$:
\begin{equation}
\left[T \chi(T)\right]_{T=0} =
{x \over 2} {S(S+1) \over 3}
= {x \over 3}
.
\label{eq:cnste}
\end{equation}
\item[(ii)] In the presence of a uniform
staggered molecular field, the susceptibility
is exponentially small at low temperature:
\begin{equation}
T \chi_{\parallel}(T) \sim \exp{\left(
- {h_s \over T} \right)}
.
\label{eq:chi-para}
\end{equation}
if the applied magnetic field is parallel
to the molecular field.  If the applied
magnetic field is perpendicular
to the molecular field, the susceptibility
tends to a constant:
\begin{equation}
T \chi(T) \sim \chi_0 T
.
\label{eq:chi-perp}
\end{equation}
If the sample is a powder as in the experiments
presented in section~\ref{sec:experiments},
the susceptibility given by
Eqs.(\ref{eq:chi-para}), (\ref{eq:chi-perp})
should be averaged over all directions
and we obtain
\begin{equation}
T \chi_{\rm av}(T) \sim \lambda\chi_0 T
\label{eq:linear}
,
\end{equation}
where $\lambda$ is smaller than unity.
\end{itemize}

Therefore a relevant question for experiments
is to determine whether $T \chi(T)$ tends
to a constant (like in Eq.~(\ref{eq:cnste}))
or tends to zero (like in Eq.~(\ref{eq:linear}))
in the limit
of zero temperature. From the answer to this
question we can determine whether 3D antiferromagnetic
correlations are important in 
Y$_2$BaNi$_{1-x}$Zn$_x$O$_5$.

\section{Experiments}
\label{sec:experiments}

\begin{table}
\begin{center}
\begin{tabular}{|@{}c@{}||@{}c@{}|@{}c@{}||@{}c@{}|@{}c@{}||@{}c@{}|@{}c@{}||@{}c@{}|@{}c@{}||@{}c@{}|@{}c@{}|}
\cline{1-11}
{ }$x$ { }&{ } $\Theta$  { }&
{ }$C$  { }& 
{ }$T_{\rm imp}$  { }& { } $C_{\rm imp}$ { } &  
{ } $\Delta_1$ { } & { } $C_1$ { } &
{ } $\Delta_2$ { } & $C_2$ & $T_c$ & $A$ \\
{ } & (K) & ($10^{-3}$ eum.K &
{ } (K) { } &  (eum.K  & 
{ } (K) { }& (eum.K & (K) & (eum.K & (K) & ($10^{-3}$eum\\
{ } & { } & {/Ba.mol)} & { } & {/Ba.mol)} & { } & {/Ba.mol)} & { } & {/Ba.mol)}
&   &{/Ba.mol)} \\ 
\cline{1-11} \cline{1-11}
{ } 0.00 { } & { } 0.222 { } & 6.44 & 71.6 & 0.017 &
534 & 1.07 &  138 &  0.267 & { } 0.021 { }  & 3.20 \\
\cline{1-11}
0.04 & 0.498 & 26.0 & 3.25  & 0.032 &
451 &  0.995 & 116 & 0.19 & 0.046 & 13.2 \\
\cline{1-11}
0.06  & 0.677 & 34.4  & 3.13  & 0.040 &
494 & 1.01 & 122 & 0.207 & 0.052 & 15.9\\
\cline{1-11}
0.08  & 0.922 & 41.6  & 3.71 & 0.046 &
486 &  0.965 & 120 & 0.19 & 0.054  &  17.2\\
\cline{1-11}
\end{tabular}
\caption{Parameters of the best fits of the susceptibility
data. $\Theta$ and $C$ are the parameters of the
Curie-Weiss fit~(\ref{eq:curie-weiss}) in the
temperature range [1~K, 10~K].
$T_{\rm imp}$ and $C_{\rm imp}$ are the parameters
of the power-law fit~(\ref{eq:power-law}) of the ``impurity'' contribution
to the susceptibility in the temperature
range [1~K, 300~K]. $\Delta_1$, $C_1$, $\Delta_2$
and $C_2$ are the parameters of the two-exponential
fit of the Haldane gap susceptibility
(see Eq.~(\ref{eq:2exp})).
$T_c$ and $A$ are the parameters of the fit 
given by Eq.~(\ref{eq:log}) in the temperature
range [100~mK, 1~K].
}
\label{table}
\end{center}
\end{table}

\begin{figure}[thb]
\centerline{\fig{7cm}{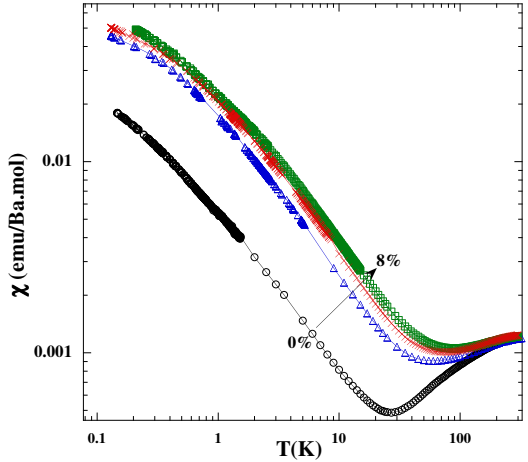}} 
\medskip
\caption{Temperature dependence of the
dc-susceptibility of the four samples 
($x=0.00$, $x=0.04$, $x=0.06$, $x=0.08$)
with a magnetic field $H=0.1$~T. This is a log-log plot.}
\label{fig:XvsT}
\end{figure}
\begin{figure}[thb]
\centerline{\fig{7cm}{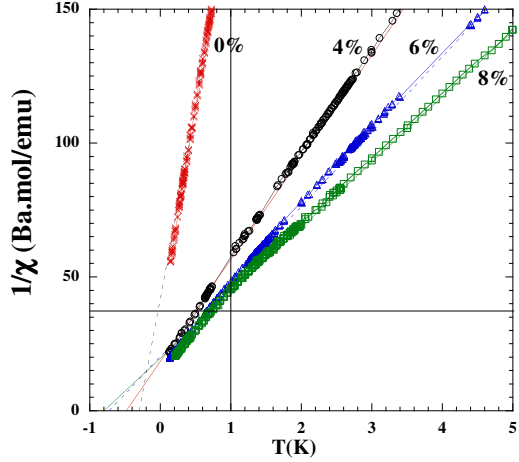}} 
\medskip
\caption{$1/\chi$ {\sl versus} $T$
for $x=0.00$, $x=0.04$, $x=0.06$ and $x=0.08$.
The dashed line are linear fits valid in
the temperature range [1~K, 10~K].
The fits intercept the $T$ axis at a negative temperature
which is a signature of antiferromagnetic correlations.} 
\label{fig:1_XvsT}
\end{figure}
\begin{figure}[thb]
\centerline{\fig{7cm}{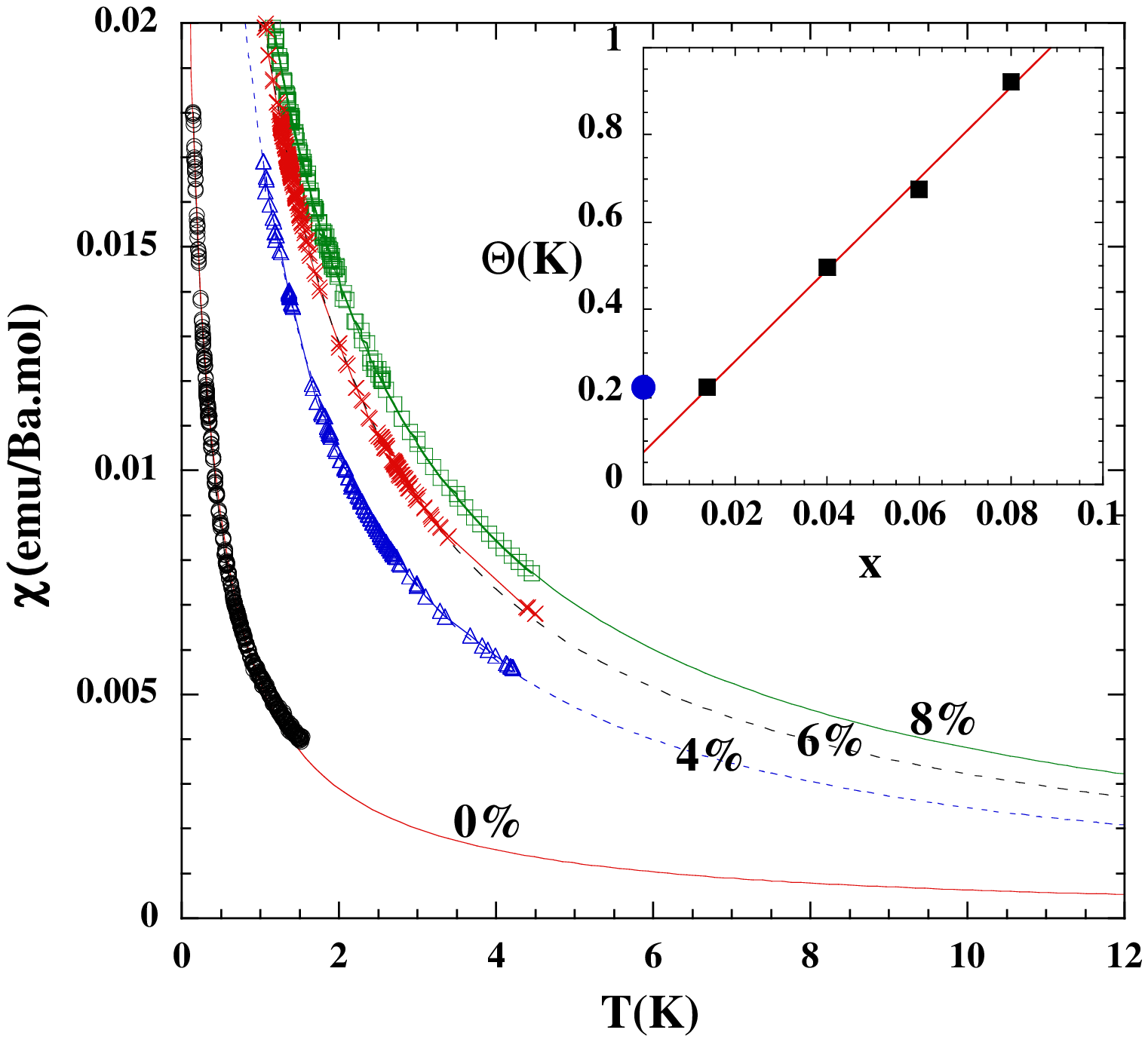}} 
\medskip
\caption{Magnetic susceptibility {\sl versus} temperature
at low temperature
for $x=0.00$, $x=0.04$, $x=0.06$ and $x=0.08$.
The  curves have been fitted to 
the Curie-Weiss law~(\ref{curie-weiss})
over the temperature range
[1~K, 10~K]. The
insert shows how $\Theta$ depends on the doping
concentration $x$. To incorporate the existence
of a finite level of disorder in the undoped
sample, we made the replacement
$x=0.00 \rightarrow x=0.014$.
}
\label{fig:Curie}
\end{figure}

We investigated the very low temperature susceptibility of 
the doped Haldane gap compound Y$_{2}$BaNi$_{1-x}$Zn$_{x}$O$_5$.
We have 
performed ac- and
dc-susceptibility measurements with a SQUID 
magnetometer equipped with a dilution refrigerator capable of 
measuring down to $100$~mK and in magnetic fields
up to $8$~Tesla. The samples are the same as those
used in
Ref.~\cite{Payen00}. In particular,
there is one undoped sample with
$x=0.00$ and three samples with
a Zn doping of $x=0.04$, $x=0.06$ and $x=0.08$.

Fig. \ref{fig:XvsT} shows
the temperature dependence of the dc-susceptibility
in a log-log plot
for the four samples where the low temperature
data from $100$~mK to $6$~K are new, and the
high temperature data from 2.4~K to 300~K
correspond to Ref.~\cite{Payen00}.
Both measurements were taken in a static
field of $0.1$~Tesla. As can be
seen, the data agree very well and no
adjustments have been made in the overlap
temperature interval.

From Fig. \ref{fig:1_XvsT} showing $1/\chi(T)$
{\sl versus} $T$,
we conclude that our experiments suggest the 
existence of {\sl antiferromagnetic correlations at very
low temperature} because the fits 
intercept the $T$ axis at a negative temperature $\Theta$.

The susceptibility in the temperature interval
[1~K, 10~K] can be more or less described by
a Curie-Weiss law
\begin{equation}
\chi(T) =\frac{C}{T+\Theta} 
\label{eq:curie-weiss}
\label{curie-weiss}
.
\end{equation}
From the insert of Fig.~\ref{fig:Curie} showing $\Theta$
{\sl versus} $x$, we deduce the following:
\begin{itemize}
\item[(i)] $\Theta$ does not extrapolate to $0$
when $x=0.00$. This is due to the
fact that the pure sample contains already a
certain level of disorder
(see Ref.~\cite{Payen00}).
We can use $x=0.014$ to describe the 
pure sample (see Fig.~\ref{fig:Curie}).

\item[(ii)] $\Theta$ increases with $x$ which
means that the strength of antiferromagnetic
correlations increases with the doping concentration.

\item[(iii)] The $\Theta$ {\sl versus} $x$
variation is similar to Cu$_{1-x}$Zn$_x$GeO$_3$
(see Refs.~\cite{Manabe98,Melin-SP}). 
The similitude between Cu$_{1-x}$Zn$_x$GeO$_3$
and Y$_2$BaNi$_{1-x}$Zn$_x$O$_5$
is another indication in favor
of the establishment of 3D antiferromagnetic correlations
in Y$_2$BaNi$_{1-x}$Zn$_x$O$_5$.
\end{itemize}

\begin{figure}[thb]
\centerline{\fig{7cm}{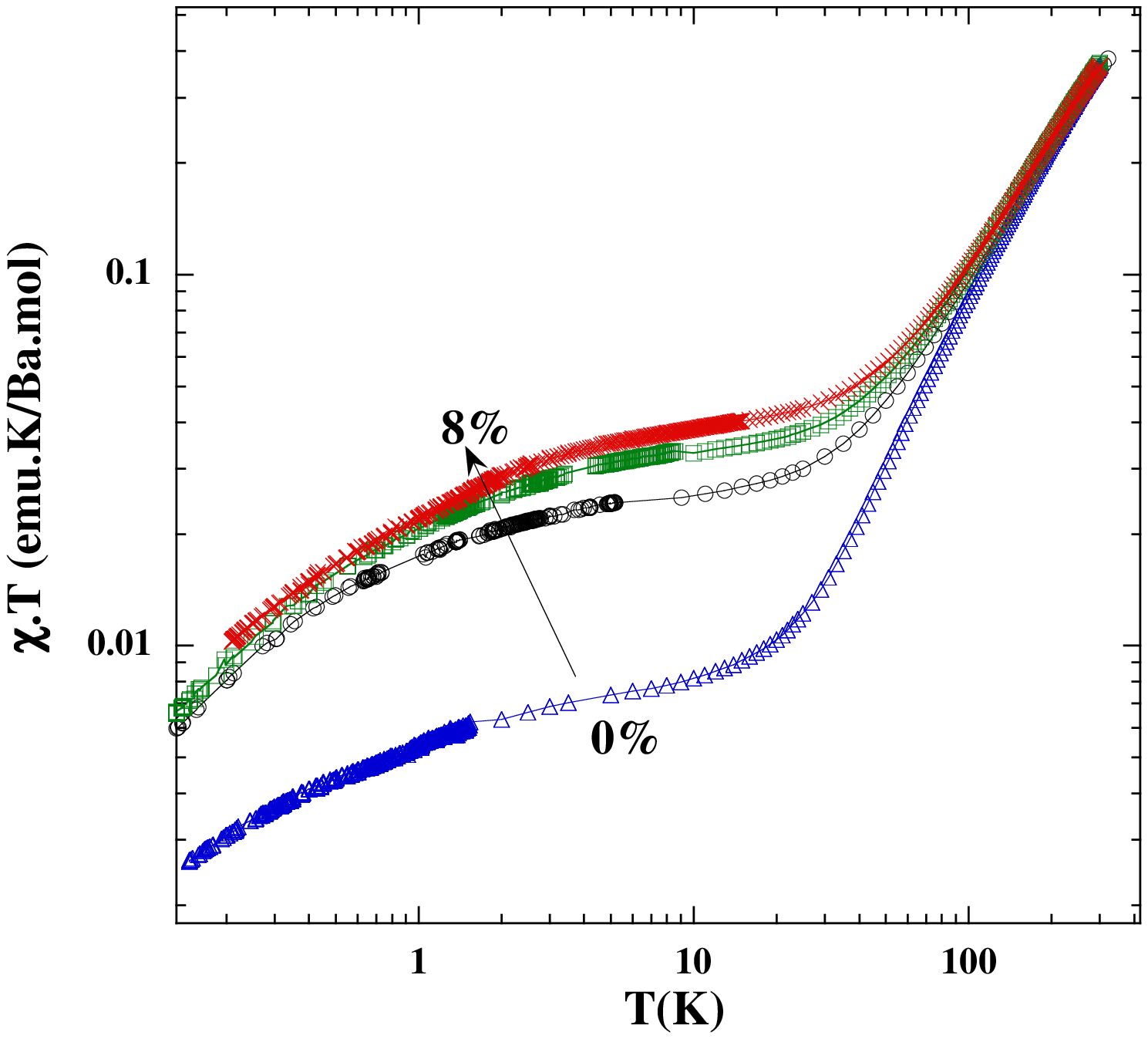}} 
\medskip
\caption{Fit of the experimental susceptibility
in the temperature range [100~mK, 300~K].
We have shown
$T \chi(T)$ {\sl versus} T in a log-log plot.
The susceptibility in the temperature
range [100~mK, 1~K] has been fitted to
an exponential term
$C_3 \exp{\left( - \frac{\Delta_3}{T} \right)}$.}
The Haldane gap contribution~(\ref{eq:2exp})
is extremely weak at low temperature.
For instance one has $T \chi_{\rm Hald}(T)=
1.2 \times 10^{-6}$emu.K/Ba.mol if
$T=10$~K.
\label{fig:XTvsT}
\end{figure}

Let us now improve the analysis of the low
temperature susceptibility.
We found that the experimental data 
in the temperature range [100~mK, 300~K] can be
well described
by a sum of two contributions (see Fig.~\ref{fig:XTvsT}):
\begin{equation}
  T \chi (T) = T \chi_{\rm imp}(T)
  + T \chi_{\rm Hald}(T)
  \label{eq:fit-delta}
  ,
\end{equation}
where $\chi_{\rm Hald}(T)$ is independent of doping
concentration
and describes the susceptibility
associated with the formation of the Haldane gap
at high temperature. $\chi_{\rm imp}(T)$ describes
the low temperature susceptibility associated
with the introduction of doping.
Following a suggestion by Souletie {\sl et al.}~\cite{Souletie01}
we have attempted to represent $T \chi_{\rm Hald}(T)$
by the sum of two exponential contributions
\begin{equation}
\label{eq:2exp}
T \chi_{\rm Hald}(T) = C_1 \exp(-\frac{\Delta_1}{T})
+ C_2 \exp(-\frac{\Delta_2}{T})
,
\end{equation}
rather than the usual activated behavior
\begin{equation}
\label{eq:magnon}
\chi_{\rm Hald}(T) = A T^{-1/2}
\exp{ \left( - \Delta_{\rm Hald}/T\right)}
\end{equation}
based on a quadratic expansion of the magnon dispersion
relation around the minimum at $q=\pi$~\cite{Regnault}.
The form~(\ref{eq:magnon}) of the susceptibility is
a good description of Haldane gap chains
at temperatures smaller than
$\Delta_{\rm Hald}/2$~\cite{Takigawa,Kim}.
At higher temperatures,
Souletie {\sl et al.}~\cite{Souletie01} have
found that~(\ref{eq:2exp}) accounts extremely well
for the results of exact diagonalizations
and DMRG calculations on finite rings with
an even number $N$ of spin-$1$ Heisenberg spins,
extrapolated to the limit $N \rightarrow
+ \infty$. The agreement between these numerical
calculations and Y$_2$BaNi$_{1-x}$Zn$_x$O$_5$
is even quantitative
since we found $\Delta_2 / \Delta_1 \simeq 3.8$
in Y$_2$BaNi$_{1-x}$Zn$_x$O$_5$,
instead of $\Delta_2 / \Delta_1 \simeq 3.86$
from the numerical calculations, and
$C_2 / C_1 \simeq 0.25$ instead
of $C_2/C_1 \simeq 0.32$.
Unlike Eq.~(\ref{eq:magnon}),
the expression (\ref{eq:2exp})
tends to the Curie law
in the high temperature limit. 
It is particularly interesting that the
expression~(\ref{eq:2exp})  reaches a physically
sound high temperature limit because
this is in this limit that the Haldane
term constitutes most of the magnetic
signal. We took advantage
of this situation to evaluate
the Haldane gap term with the data
above $1$~K and subtracted the corresponding
contribution to extract an improved
estimate of the impurity contribution.
\begin{figure}[thb]
\centerline{\fig{7cm}{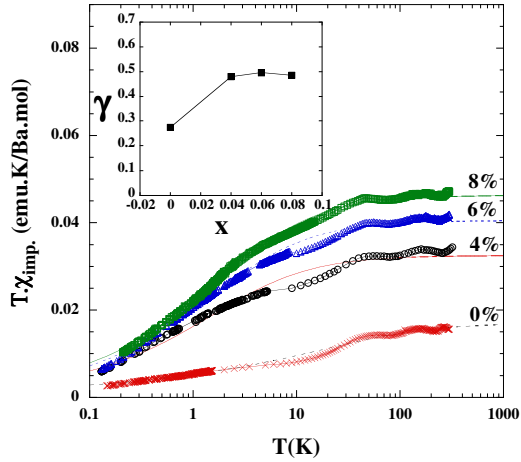}} 
\medskip
\caption{
  Temperature dependence of $T \chi_{\rm imp}(T)$.
  The fits correspond to (\ref{eq:power-law}).
  The insert show how $\gamma$
  depends on the doping concentration $x$.
}
\label{fig:Xtdiff}
\end{figure}
The difference $\chi_{\rm imp}(T)
= \chi(T) - \chi_{\rm Hald}(T)$ can be interpreted
as the susceptibility of the spin-$1/2$ moments
generated by the Zn ions. We have shown on
Fig.~\ref{fig:Xtdiff} the temperature dependence
of $T \chi_{\rm imp}(T)$.
We verified that the high temperature limit
of $T \chi_{\rm imp}(T)$ corresponds approximately to
a concentration $2x$ of spin-$1/2$
moments. 

\begin{figure}[thb]
\centerline{\fig{8cm}{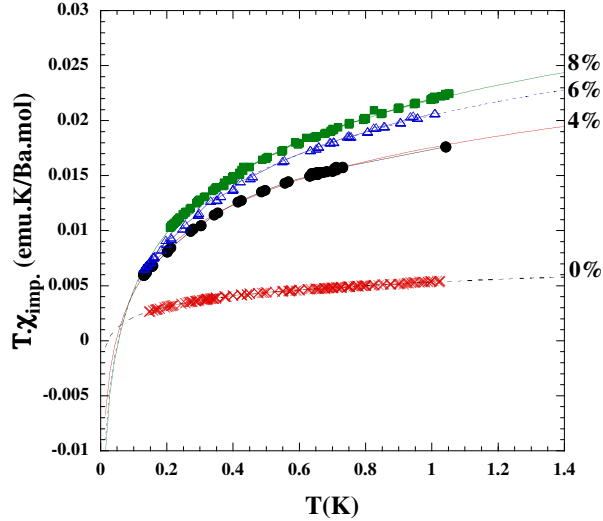}} 
\medskip
\caption{
  Fit of $T \chi_{\rm imp}(T)$ to
  the form~(\ref{eq:ln}) in the temperature
  range [100~mK, 1~K].
}
\label{fig:log}
\end{figure}

To describe the variation of $T \chi_{\rm imp}(T)$ we
use two different expressions in the
two temperature intervals [100~mK, 1~K]
and [1~K, 300~K].
The impurity contribution shown in
Fig.~\ref{fig:Xtdiff}
has been fitted by a power law
in the temperature range [1~K, 300~K] :
\begin {equation}
  T \chi_{\rm imp}(T) =  C_{\rm imp}
  \left(1+\frac{T_{\rm imp}}{T}\right)^{-\gamma}
  \label{eq:power-law}
  ,
\end{equation}  
where $C_{\rm imp}$ and $T_{\rm imp}$ are given
in Table~\ref{table}, and $\gamma$ is shown
on the insert in Fig.~\ref{fig:Xtdiff}.
We have found for the
temperature range [100~mK, 1~K] that the experimental
data are better described by a logarithmic
temperature dependence (see Fig.~\ref{fig:log}):
\begin{equation}
\label{eq:ln}
\label{eq:log}
T \chi_{\rm imp}(T) = A \log_{10}{\left({T \over T_c } \right)}
,
\end{equation}
where $A$ and $T_c$ are given in Table~\ref{table}.
Note that in this temperature range the contribution
from the Haldane gap susceptibility~(\ref{eq:2exp})
is negligible.

From the two fits given by~(\ref{eq:power-law})
and~(\ref{eq:ln}) we deduce that
$T \chi(T)$ tends to zero in the limit 
$T \rightarrow 0$. Regarding the discussion in
section~\ref{sec:RG-chi} we deduce that interchain
interactions play an important role at low temperature.
This validates the molecular field approach used
in our theoretical description.
Moreover the fit~(\ref{eq:power-law}) suggests the
existence of a finite N\'eel temperature $T_c$.
We notice that
\begin{itemize}
\item[(i)] The value of $T_c$ (see Table~\ref{table})
is too small to be probed directly in experiments.
\item[(ii)] $T_c$ deduced from the fit~(\ref{eq:ln})
is one order of magnitude
smaller than $\Theta$ deduced from the
Curie-Weiss fit~(\ref{eq:curie-weiss}).
\item[(iii)] $T_c$ increases with $x$, which
is compatible with an interpretation in terms
of a N\'eel temperature.
\end{itemize}
\begin{figure}[thb]
\centerline{\fig{10cm}{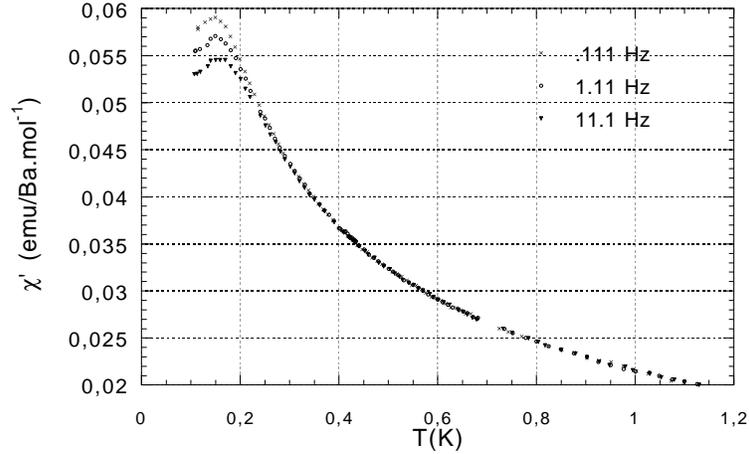}} 
\medskip
\caption{Temperature dependence of the ac-susceptibility
$\chi'(T)$
with $x=0.08$ for frequencies
$f=0.11$~Hz, $f=1.1$~Hz, $f=11.1$~Hz.
The temperature of the
maximum of
$\chi'(T)$ is almost frequency-independent.
}
\label{fig:ac}
\end{figure}

Finally we also performed ac-susceptibility measurements
which are shown on Fig.~\ref{fig:ac} for three
different frequencies. We have obtained a maximum
in $\chi'(T)$ at a temperature $T\simeq 160$~mK.
The temperature of the maximum in the
real part of the susceptibility is
frequency-independent. This indicates that
in spite of a slowing down
of the dynamics at very low temperature,
Y$_2$BaNi$_{1-x}$Zn$_x$O$_5$ is not
a spin glass at very low temperature.
This may be contrasted with the hole doped compound
Y$_{2-x}$Ca$_x$BaNiO$_5$~\cite{Payen01} (with
$0 \le x \le 0.2$) in
which a spin glass thermodynamic transition
has been reported at $T_g=2 -3$~K.
The very low temperature ac-susceptibility
of Y$_2$BaNi$_{1-x}$Zn$_x$O$_5$ represented on
Fig.~\ref{fig:ac}
is reminiscent of the dc-susceptibility
of the mean field model (see Fig.~\ref{fig:chi-2spin}).
We suggest that the very low temperature
ac- and dc-susceptibilities (see Figs.~\ref{fig:log}
and~\ref{fig:ac})
can be interpreted in terms of antiferromagnetic
correlations at low temperature.

\section{Conclusion}
\label{sec:conclusion}

To conclude, we have presented a detailed study of
antiferromagnetic correlations in
Y$_2$BaNi$_{1-x}$Zn$_x$O$_5$. Based on the quantum
chemistry calculation in Ref.~\cite{Chemist},
we have proposed 
that Y$_2$BaNi$_{1-x}$Zn$_x$O$_5$
can be well described by a model
in which interchain couplings are treated 
in mean field. The ground state of this model
shows a coexistence between random dimers
and antiferromagnetism. In the presence of the
staggered molecular field, the
product $T\chi(T)$ goes to zero in the limit
of zero temperature.

We have presented very low susceptibility experiments
and shown how to subtract from the total
susceptibility the high energy contribution 
corresponding to the formation of the Haldane
gap. The product
$T \chi_{\rm imp}(T)$ tends to zero at
low temperature. In view of the model, we conclude
that this behavior is due to interchain correlations.
At very low temperature the product $T \chi_{\rm imp}(T)$
has a logarithmic temperature dependence
in the temperature interval [100~mK, 1~K]:
$T \chi_{\rm imp}(T) \simeq A \log_{10}{(T/T_c)}$.
If we take for granted that this scaling
is valid at lower temperatures we deduce
the existence of a finite N\'eel transition
temperature, which is in agreement
with the assumption made in the mean field
model. We have also obtained a maximum
in the temperature dependence of the
ac-susceptibility $\chi'(T)$ which 
is an indication of the presence
of antiferromagnetic correlations at very
low temperature.

An important question left open is to clarify
the behavior of 2D and 3D models and determine
whether there is an antiferromagnetic phase transition
at a finite temperature. Using a mean field model
we have shown here that there is a coexistence between
dimer formation (being due mainly to the even
segments coupled antiferromagnetically) and
antiferromagnetism (being due mainly to odd segments
coupled ferromagnetically). If interchain interactions
are very small, it is possible that singlets
can be formed also among neighboring chains. These
singlets would compete with 3D antiferromagnetism. An
open question is to determine whether 
3D antiferromagnetism is stable against
dimer formation even with very small interchain interactions.
Answering this question may
clarify why $T_c$ (obtained from the fit~(\ref{eq:log})
in the temperature range [100~mK, 1~K]) is
one order of magnitude smaller than
$\Theta$ (obtained from the Curie-Weiss
fit~(\ref{eq:curie-weiss}) in the temperature
range [1~K, 10~K]).

\appendix

\section{Exact treatment of the two-spin model}
\label{app:exact}
\subsection{Even segments}

The Hamiltonian of even segments coupled
antiferromagnetically reads
\begin{equation}
\label{eq:H-odd}
{\cal H} = J(l) {\bf S}_1 . {\bf S}_2
- h_1 S_1^z + h_2 S_2^z
,
\end{equation}
where we included a
staggered molecular field $h_s = (h_1 + h_2)/2$
and a uniform field  $h_u = (h_1 - h_2)/2$.
The uniform field is used to calculate the uniform
susceptibility. There are two eigenstates
$|+,+ \RA$ and $|-,- \RA$ with an energy
$E_{|+,+\RA} = J/4 - h_u$, and
$E_{|-,-\RA} = J/4 + h_u$.
There are two other eigenstates
$|\psi_\epsilon \RA = x_\epsilon |+,- \RA
+ y_\epsilon |-,+ \RA$ corresponding to
$\epsilon=\pm 1$, with an energy
$$
E_\epsilon = - \frac{J}{4}
+ \frac{\epsilon}{2} \sqrt{ J^2 + 4 h_s^2}
,
$$
and
\begin{eqnarray}
\label{eq:x-eps1}
x_\epsilon &=& \sqrt{ \frac{1}{2}
- \epsilon \frac{ h_s / J}{
\sqrt{1 + 4 h_s^2 / J^2}}}
,\\
y_\epsilon &=& \epsilon \sqrt{ \frac{1}{2}
+ \epsilon \frac{ h_s / J}{
\sqrt{1 + 4 h_s^2 / J^2}}}
\label{eq:x-eps2}
.
\end{eqnarray}
The magnetization
is given by the thermal average
\begin{equation}
\label{eq:magne}
M(l,T) = \frac{ e^{-\beta E_{|+,+\RA}}
- e^{-\beta E_{|-,-\RA}}}{Z(T)}
,
\end{equation}
where the partition function is
$Z(T) = \mbox{Tr} \exp{(-\beta {\cal H})}$,
with $\beta=1/T$ the inverse temperature.

\subsection{Odd segments}
The odd segments are coupled ferromagnetically:
\begin{equation}
\label{eq:H-ferro}
\label{eq:H-even}
{\cal H} = - |J(l)| {\bf S}_1 . {\bf S}_2
- h_1 S_1^z - h_2 S_2^z
,
\end{equation}
where $h_1=h_2=h_s+h_u$ in sublattice A
and $h_1=h_2=-h_s+h_u$ in sublattice B.
The solution of (\ref{eq:H-ferro}) is straightforward.

\section{Renormalization of the two-spin model}
\label{app:RG}
We should distinguish between odd segments
and even segments.
We use the RG transformations
given in Appendices~\ref{app:RG-ferro}
and~\ref{app:RG-antiferro}.

\subsection{Even segments}
We start from the Hamiltonian given
by Eq.~\ref{eq:H-odd} in which the exchange
is antiferromagnetic, and where we assume
that there is no uniform magnetic field.
There are three energy scales in the problem:
\begin{itemize}
\item[(i)] The temperature $T$.
\item[(ii)] The exchange gap
$\Delta_J=J=\Delta \exp{(-l/\xi)}$.
\item[(iii)] The staggered molecular field gap $\Delta_h=h_s$.
\end{itemize}
There are two
relevant length scales:
\begin{itemize}
\item[(i)] The thermal length $\xi_T$
for which $\Delta_J=T$. One has
$\xi_T = \xi \ln{(\Delta/T)}$.
\item[(ii)] The staggered molecular field length for
which $\Delta_J=\Delta_h$. One has
$\xi_h = \xi \ln{(\Delta/h_s)}$.
\end{itemize}
We start the RG at high temperature,
{\sl i.e.} $T \sim \Delta$, and decrease the temperature
until $\xi_T$ becomes equal to $\xi_h$.
All the pairs of spins such that
$l < \xi_T$ are frozen into dimers.
Hence,
the dimer order parameter $\varphi_{\rm dim}$ is given by
$$
\varphi_{\rm dim}(T) =
\int_0^{\xi_T} \frac{1}{2} {\cal P}(l) dl
= \frac{1}{2} \left[ 1 - \left( \frac{T}{\Delta}
\right)^{x \xi} \right]
\mbox{, where $T>h_s$}
.
$$
When the temperature becomes smaller
than the staggered molecular
field $h_s$, the spins that have not yet been decimated
({\sl i.e.} having $l > \xi_h$) contribute
to the antiferromagnetic order parameter:
$$
\varphi_{\rm AF}(T) =
\int_{\xi_h}^{+ \infty} \frac{1}{2}
{\cal P}(l) dl = \frac{1}{2}
\left( \frac{h_s}{\Delta} \right)^{x \xi}
\mbox{, where $T < h_s$}
.
$$

\subsection{Odd segments}
The odd segments are coupled ferromagnetically,
with the Hamiltonian Eq.~\ref{eq:H-ferro}.
At high temperature,
the RG transforms the pairs
of spin-$1/2$ moments having $l<\xi_T$
into spin-$1$ moments, having a renormalized
staggered molecular field equal to
$\tilde{h}_s=(h_1+h_2)/2=h_s$.
The density of spin-$1$ moment reads
$$
\varphi_1(T) = \int_0^{\xi_T} \frac{1}{2}
{\cal P}(l) dl = \frac{1}{2}
\left[ 1 - \left( \frac{T}{\Delta} \right)^{x \xi}
\right] \mbox{, where $T > 2 h_s$}
.
$$
These spin-$1$ moments have a gap
$\Delta_{h}=2 h_s$. When
$T=2 h_s$, all of the spin-$1$ moments
are frozen antiferromagnetically.
The antiferromagnetic order parameter
jumps from $\varphi_{\rm AF}(2 h_s + 0^+)=0$
to
$$
\varphi_{\rm AF}(2 h_s - 0^+) =
\frac{1}{2}
\left[ 1 - \left( \frac{2 h_s}{\Delta}
\right)^{x \xi} \right]
.
$$
When $h_s < T < 2 h_s$, the pairs
of spin-$1/2$ moments having a length
$\xi_{2 h_s} < l < \xi_T$
are transformed into spin-$1$
objects, which are immediately frozen
antiferromagnetically. The 
antiferromagnetic order parameter is
$$
\varphi_{\rm AF}(T) = \varphi_{\rm
AF}(2 h_s - 0^+) + \int_{\xi_{2 h_s}}^{\xi_T}
\frac{1}{2} {\cal P}(l) dl
= \frac{1}{2} \left[1 - \left(
\frac{T}{\Delta} \right)^{x \xi} \right]
\mbox{, where $h_s < T < 2 h_s$}
.
$$
When $T=h_s$, the survival spin-$1/2$ moments are
directly
frozen antiferromagnetically. The resulting
antiferromagnetic order parameter is
$$
\varphi_{\rm AF}(T) = \varphi_{\rm AF}(h_s) +
\int_{\xi_{h_s}}^{+ \infty} \frac{1}{2}
{\cal P}(l) dl = \frac{1}{2} 
\mbox{, where $T<h_s$}
.
$$

\section{RG transformations of ferromagnetic bonds}
\label{app:RG-ferro}

Let us consider the Hamiltonian
\begin{equation}
\label{eq:Hamil-F}
{\cal H} = - J_{1,2} {\bf S}_1 . {\bf S}_2
-h_1 S_1^z - h_2 S_2^z
.
\end{equation}
in which the spins $S_1$ and $S_2$ are
coupled ferromagnetically.
We replace the two spins $S_1$ and $S_2$
by an effective spin $S=S_1+S_2$ with a Hamiltonian
${\cal H}=- \tilde{h} S^z$, and
need to calculate the
renormalized field $\tilde{h}$. We use
two different methods giving the same
answer in the ferromagnetic case: (i)
a semi-classical calculation using Holstein-Primakov
bosons;
(ii) a quantum mechanical
calculation using Clebsch-Gordon coefficients.

\subsection{Holstein-Primakov bosons}
We replace the spin operators by their bosonic
representation $S_1^z=S_1 - a^+ a$,
$S_1^+ = \sqrt{2 S_1} a$, $S_1^-=\sqrt{2 S_1} a^+$,
and $S_2^z=S_2 - b^+ b$,
$S_2^+ = \sqrt{2 S_2} b$, $S_2^-=\sqrt{2 S_2} b^+$.
The Hamiltonian Eq.~\ref{eq:Hamil-F} becomes
\begin{equation}
\label{eq:boson-F}
{\cal H} = \left( - J S_2 + h_1 \right) a^+ a
+ \left( -J S_1 + h_2 \right) b^+ b
+ J \sqrt{S_1 S_2} \left( ab^+ + a^+ b \right)
,
\end{equation}
where we discarded the classical energy contribution.
The Hamiltonian Eq.~\ref{eq:boson-F}
is diagonalized by the unitary transformation
$a=c \cos{\theta} + d \sin{\theta}$, and
$b=-c \sin{\theta} + d \cos{\theta}$, where
$$
\tan{(2 \theta)} =
- \frac{ 2 J \sqrt{S_1 S_2}}{
J(S_1-S_2) + h_1 - h_2}
.
$$
Expanding the low energy Hamiltonian
to lowest order in $h_1$ and $h_2$
leads to
\begin{equation}
\label{eq:RG-ferro}
\tilde{h}(S_1 + S_2) = h_1 S_1
+ h_2 S_2
.
\end{equation}

\subsection{Clebsch-Gordan coefficients}
We combine the two spins $S_1$ and $S_2$
to form a state of the type
$|S_1,S_2 | S_1^z,S_2^z \RA$. The 
highest weight state is
$|S,S \RA = |S_1,S_2| S_1,S_2 \RA$,
and we need to determine the states
$|S,M \RA$ to calculate
$\LA S,M | - h_1 S_1^z - h_2 S_2^z
| S,M \RA$.

It is straightforward to show that
\begin{equation}
\label{eq:Sz=S}
\LA S,S | - h_1 S_1^z - h_2 S_2^z
| S,S \RA = -h_1 S_1 - h_2 S_2
.
\end{equation}
Identifying this with
$-\tilde{h} S$, we obtain the renormalized
magnetic field Eq.~\ref{eq:RG-ferro}.
One can check that Eq.~\ref{eq:RG-ferro}
is also valid for spin-$S$ states having
lower values of $S^z$. For instance,
one finds
$$
|S,S-1 \RA = \sqrt{\frac{S_1}{S}}
|S_1,S-2 | S_1-1,S_2 \RA +
\sqrt{ \frac{S_2}{S}} |S_1,S_2 | S_1,S_2-1 \RA
,
$$
leading to
\begin{equation}
\label{eq:state-1}
\LA S,S-1 | - h_1 S_1^z - h_2 S_2^z | S,S-1 \RA
= -\frac{S-1}{S} \left(h_1 S_1 +h_2 S_2 \right)
.
\end{equation}
Eq.~\ref{eq:state-1} implies
directly Eq.~\ref{eq:RG-ferro}.

It can also be shown that
\begin{eqnarray}
|S,S-2 \RA &=& \sqrt{ \frac{ S_1 ( 2 S_1-1)}
{S(2S-1)}} |S_1,S_2 | S_1-1,S_2 \RA\\
&+& 2 \sqrt{ \frac{S_1 S_2}{S(2S-1)}}
|S_1,S_2 | S_1-1,S_2-1 \RA\\
&+& \sqrt{ \frac{S_2 ( 2 S_2-1)}{S(2S-1)}}
|S_1,S_2 | S_1,S_2-2 \RA
.
\end{eqnarray}
We deduce
\begin{equation}
\label{eq:state-2}
\LA S,S-2 | - h_1 S_1^z - h_2 S_2^z | S,S-2 \RA
= -\frac{S-2}{S} \left(h_1 S_1 +h_2 S_2 \right)
,
\end{equation}
leading again to Eq.~\ref{eq:RG-ferro}.

We are lead to conjecture
that 
\begin{equation}
\label{eq:renor-gen-F}
\LA S,M| -h_1 S_1^z - h_2 S_2^z | S,M \RA
= - \frac{M}{S} \left( h_1 S_1 +h_2 S_2 \right)
\end{equation}
for any value of $M$. 

\section{RG transformations of antiferromagnetic bonds}
\label{app:RG-antiferro}

Let us consider the Hamiltonian
\begin{equation}
\label{eq:H-app-anti}
{\cal H} =  J_{1,2} {\bf S}_1 . {\bf S}_2
-h_1 S_1^z - h_2 S_2^z
\end{equation}
in which the spins $S_1$ and $S_2$ are
coupled antiferromagnetically.
We assume that $S_1 > S_2$ and look
for the effective Hamiltonian under the form
$- \tilde{h} S^z$, with
$S=S_1-S_2$.

\subsection{Holstein-Primakov bosons}
To obtain the RG equation in the semiclassical
limit, we represent the two spins by
Holstein-Primokov bosons:
$S_1^z=S_1 - a^+ a$,
$S_1^+=\sqrt{2 S_1} a$,
$S_1^-=\sqrt{2 S_1} a^+$, and
$S_2^z=-S_2+b^+ b$,
$S_2^+ = \sqrt{2 S_2} b^+$,
$S_2^-=\sqrt{2 S_2} b$.
The Hamiltonian Eq.~\ref{eq:H-app-anti} becomes
\begin{equation}
\label{eq:bos}
{\cal H} = (J S_2 + h_1)a^+ a
+ (J S_1-h_2) b^+ b
+ J \sqrt{S_1 S_2}
(a b + a^+ b^+ )
,
\end{equation}
where we retained only the quadratic contributions,
and discarded the classical energy term.
The bosonic Hamiltonian Eq.~\ref{eq:bos}
is readily diagonalized by the Bogoliubov
rotation $a = c \cosh{\theta} 
+ d^+ \sinh{\theta}$, with
$$
\tanh{(2 \theta)} =
- \frac{ 2 J \sqrt{S_1 S_2}}
{ J(S_1 + S_2) + h_1 - h_2}
.
$$
Expanding the low energy Hamiltonian
to lowest order in $h_1$ and $h_2$
leads to
\begin{equation}
\label{eq:renor-cl}
\tilde{h} (S_1 - S_2)  = h_1 S_1 - h_2 S_2
,
\end{equation}
which constitutes the {\sl classical} renormalization
equation. As we show in sections~\ref{sec:AF-CJ-1/2}
and~\ref{sec:AF-CJ-1},
the {\sl quantum} renormalization equation
does not coincide exactly with the classical
equation.

\subsection{Clebsch-Gordan coefficients: coupling of a
spin-$s$ to a spin-$1/2$}
\label{sec:AF-CJ-1/2}

Considering $S_1=s>1/2$ and $S_2=1/2$, we find
$$
|s-1/2,M \RA = \sqrt{ \frac{s+M+1/2}{2s+1}}
|s,1/2 | M+1/2,-1/2 \RA
- \sqrt{ \frac{s-M+1/2}{2s+1}}
|s,1/2 | M-1/2,1/2 \RA
,
$$
from what we deduce
$
\LA s-1/2,M | -h_1 S_1^z - h_2 S_2^z | s-1/2,M \RA
= - \tilde{h} M 
$, with
\begin{equation}
\label{eq:renor-1/2}
\tilde{h}
\left(s+ \frac{1}{2} \right)  =
h_1 (s+1) - \frac{1}{2} h_2
.
\end{equation}
This RG equation is clearly not identical to
Eq.~\ref{eq:renor-cl}.

\subsection{Clebsch-Gordan coefficients: coupling of a
spin-$s$ to a spin-$1$}
\label{sec:AF-CJ-1}
Now let us consider $S_1=s>1$ and $S_2=1$.
Using standard orthonormalization and recursion techniques,
we obtain
\begin{eqnarray}
|s-1,M \RA &=&
- \sqrt{ \frac{ (s-M)(s-M+1)}{2s ( 2s+1)}}
|s,1 | M-1,1 \RA \\
\nonb
&+& \sqrt{\frac{(s+M)(s-M)} {s (2s+1)}}
|s,1|M,0 \RA\\
\nonb
&-& \sqrt{ \frac{ (s+M+1)(s+M)}
{2s ( 2s+1)}}
|s,1|M+1,-1 \RA
,
\end{eqnarray}
form what we deduce
$
\LA s-1,M | -h_1 S_1^z - h_2 S_2^z
|s-1,M \RA = - \tilde{h} M
,
$
with the renormalized magnetic field
\begin{equation}
\label{eq:renor-1}
\tilde{h} s = h_1 ( s+1) - h_2
.
\end{equation}
Comparing Eqs.~\ref{eq:renor-1/2} and
Eq.~\ref{eq:renor-1}, we are lead to
conjecture that the general renormalization
equation is
\begin{equation}
\tilde{h} (S_1-S_2+1)  = 
h_1 (S_1+1) - h_2 S_2 \mbox{  (with $S_1 > S_2$)}
\label{eq:renor-qu}
.
\end{equation}
The quantum RG equation
Eq.~\ref{eq:renor-qu}
does not coincide exactly with the classical 
one (see Eq.~\ref{eq:renor-cl}). Nevertheless,
the large-$S$ limit of the quantum RG equation
Eq.~\ref{eq:renor-qu}
does coincide with 
the classical equation Eq.~\ref{eq:renor-cl}.

\end{document}